\documentclass[12pt]{article}
\usepackage{latexsym}
\usepackage[dvips]{epsfig}
\usepackage{graphpap}

\def\Journal#1#2#3#4#5#6{(#5) ``#6'' {#1} {\bf #2} #3#4}

\newcommand{\bm}[1]{\mbox{\boldmath $#1$}}

\def\CQG{\em Class. Quantum Grav.}
\def\JPA{\em J. Phys. A: Math. Gen.}
\def\PRD{\em Phys. Rev. D }
\def\GRG{\em Gen. Rel. Grav.}

\def\JMP{\em J. Math. Phys.}

\def\CMP{\em Commun. Math. Phys.}

\def\PRL{\em Phys. Rev. Lett.}

\def\AP{\em Ap. J.}

\def\PREP{\em Phys. Rep.}

\def\ANY{\em Ann. Phys. (N. Y.)}
\def\SPJ{\em Sov. Phys. JETP}

\def\PPLL{\em Phys. Lett. A }

\def\AJ{\em Ap. J.}

\def\Pary{\frac{\partial}{\partial y}}
\def\Parz{\frac{\partial}{\partial z}}
\def\Parx{\frac{\partial}{\partial x}}
\def\Part{\frac{\partial}{\partial t}}

\def\Paru{\frac{\partial}{\partial u}}
\def\Parv{\frac{\partial}{\partial v}}

\def\acc{\mbox{a}}         
\def\d{{\rm \mbox{d}}}
\def\g{{\rm \mbox{g}}}

\def\lie{{\pounds}}

\def\fuma{\left(e^{\mu t}-1\right)}
\def\fume{\left(1-e^{-\mu t}\right)}

\def\be{\begin{equation}}
\def\ee{\end{equation}}
\def\bea{\begin{eqnarray}}
\def\eea{\end{eqnarray}}
\def\bean{\begin{eqnarray*}}
\def\eean{\end{eqnarray*}}

\def\D{{\rm \mbox{D}}}

\def\kv{(\kappa_v v)}
\def\ku{(\kappa_u u)}

\begin{document}

\title{New family of inhomogeneous $\gamma$-law cosmologies: example of
gravitational waves in a homogeneous $p=\varrho/3$ background}

\author{
Jos\'e M. M. Senovilla\thanks{Also at
Laboratori de F\'{\i}sica Matem\`atica,
Societat Catalana de F\'{\i}sica, IEC, Barcelona.} \footnotemark[4]
and
Ra\"ul Vera\footnotemark[1] $^\ddag$\\
\footnotemark[4] Fisika Teorikoaren Saila, Euskal Herriko Unibertsitatea,\\
644 P.K., 48080 Bilbao, Spain.\\
\ddag School of Mathematical Sciences, Queen Mary and Westfield College,\\
Mile End Road, London E1 4NS, England.\\}

\maketitle

\begin{abstract}
We present an explicit three-parameter class of $p=\gamma \varrho$,
($-1/3\leq\gamma<1$), cosmological models admitting a two-dimensional
group $G_{2}$ of isometries acting on spacelike
surfaces. The family is self-similar in the sense that it has a further
homothetic vector field and it contains subfamilies of both 
(previously unknown) tilted and 
non-tilted Bianchi models with
that equation of state. This is the first algebraically general
class of solutions of this kind including dust inhomogeneous solutions.
The whole class presents a universal spacelike big-bang singularity
in the finite past.
More interestingly, the case $p=\varrho /3$
constitutes a new two-parameter inhomogeneous subfamily which can
be viewed as a 
Bianchi V background with a gravitational wave travelling 
orthogonally to the surfaces of transitivity of the $G_{2}$ group. 
This wave generates the {\it inhomogeneity} of the spacetime and is 
related to
the sound waves {\it tilting} the perfect fluid. It seems to be the 
first explicit exact example of a gravitational wave 
travelling along a homogeneous background that has a realistic equation of 
state $p=\varrho /3$.
\end{abstract}

\section{Introduction}
There is no need to mention that among the various methods
used for the study of spatially inhomogeneous cosmological models,
the research on {\em exact} solutions of Einstein field equations
plays a crucial role. Due to the high non-linearity
of the equations, the exact solutions are necessary for the
understanding of particular qualitative features that may constitute
a guide in the study of general situations.
Indeed, this has been the way in which many new kind of unexpected
behaviours have been found.
Of course, the exact solutions properties must be 
related and compared with the results obtained from other methods, 
such a combination usually leads to very powerful and general 
conclusions. For instance, when using the dynamical systems techniques
in Cosmology \cite{Llibre}, some special exact solutions are shown
to be asymptotic states
of general classes of models. Exact solutions can also be compared with
approximations or perturbations to check the validity of
the involved expansions \cite{KRAM}. Yet another example could be
the study of the structure and appearance of singularities,
which complements and sheds some light onto the singularity theorems
and their conclusions, see \cite{totxosing} and references therein.

The research on exact solutions is based on some
physically reasonable restrictions used to simplify the Einstein
equations. As an outstanding example, and with regard to
geometrical properties,
the existence of symmetries described by $n$-dimensional groups of motions
$G_n$ (see \cite{KRAM} and references therein)
was the first assumption
treated in a systematic Al way, see e.g. \cite{llibrePetrov,ryan,KRAM,KR},
leading to classifications of solutions as well as to fruitful
techniques for their finding. In this sense, and with the study of 
spatially inhomogeneous cosmologies in mind, an important and 
particularly fruitful line of research during the last two decades has 
been the consideration of the class of spacetimes
admitting a maximal 2-dimensional group of isometries $G_2$ acting
on spacelike surfaces. This line was somehow launched in \cite{WW1}
with a classification scheme for the particular Abelian case
of these so-called ``$G_2$ spacetimes''
based solely on the properties and relations of the Killing vector
fields. 
The classification was generalized for the non-Abelian case in 
\cite{alitesi}, see \cite{Raultesi} for a complete review.
Among the classes defined in the Abelian case, the most
simple subcase arises when there exists a family of surfaces
orthogonal to the orbits of the group (it is then said that
the group acts orthogonally transitively) and the two
Killing vectors are mutually orthogonal, which implies that they are
in fact hypersurface orthogonal, so that the metric can be cast
in diagonal form in coordinates adapted to the Killing vectors.
Focusing the attention on these diagonal $G_2$ spacetimes,
some additional assumptions have been made
in order to simplify the field equations for a perfect fluid source,
as for example the existence of additional homothetic or proper
conformal symmetries, see \cite{ali,MATHO} and references therein.
Let us recall here that there have also been general studies on
orthogonally transitive
$G_2$ cosmologies from a qualitative point of view, analyzing
the autonomous system of first-order partial differential equations
derivable from the Einstein field equations by using methods from
the theory of dynamical systems \cite{HEWWAIN,RUSE,Llibre}. The relations
between some of the known explicit solutions and these
theoretical studies were widely analyzed in \cite{Llibre} and many references
therein.

Another important simplifying assumption for the perfect-fluid 
diagonal $G_{2}$ spacetimes, which has received
systematic attention, corresponds to
the separability of the metric functions in coordinates
that keep the diagonal form of the metric, called {\em canonical} coordinates.
The case when these canonical coordinates
that bring the metric functions to a separate form
are also adapted to the velocity vector of the fluid, that is to say, 
they are {\em comoving} coordinates too, was
exhausted in \cite{WAGO,RUSE,AGGO},
except for a very particular case identified in \cite{Raultesi}
that did not appear in \cite{RUSE} accidentally.
The general treatment of the separability in non-comoving canonical coordinates
can be found in \cite{G2NC,Raultesi},
where a classification for separable diagonal $G_2$ on $S_2$
perfect-fluid solutions was obtained depending
on the number of linearly independent functions appearing in the metric,
leading to a systematic procedure for the obtaining of solutions.
The classification was exhausted, but not wholly solved,
because once the machinery for the systematic derivation of solutions
was established, the main effort was focused on finding solutions with
special interest or physical relevance in order to study them in detail.

Thus, for instance, an interesting solution was singled out
in \cite{Raultesi} (named as 22BIIc) 
because of its $\gamma$-law equation of state which includes
the relevant cases $\gamma=0$ (dust models)
and $\gamma=1/3$ (models for relativistic radiation).
It is also interesting because it provides
inhomogeneous generalizations of some Bianchi III, V and VI$_h$
models found in \cite{EM,ruban}, see also \cite{Llibre}.
The particular dust solutions belonging
to this family are actually included in one of the two classes
of dust spacetimes studied in \cite{SEVEdust}.
The solutions with non-zero $\gamma$, including those with 
$\gamma=1/3$, are new, though. In fact,
the number of exact solutions for inhomogeneous
spacetimes with a $p=\varrho /3$ equation
of state is rather scarce: as far as we know
the first one appeared in the Wainwright--Goode
family \cite{WAGO}
to which followed the Feinstein--Senovilla solution
\cite{alexjose1}, Davidson's \cite{davidson90},
the singularity-free metric of \cite{senosing}, their
common generalization in the
Ruiz--Senovilla class \cite{RUSE},
and the non-diagonal $p=\gamma \varrho$ family found
by Mars and Senovilla \cite{Marctesi,MASEbi}.

The aim of this paper is to present the explicit family of solutions 
mentioned in the previous paragraph, as well as to 
perform an extensive detailed geometrical and physical study of its main
features. The solutions constitute a
three-parameter class of $p=\gamma \varrho$,
$-1/3\leq\gamma<1$, cosmological models
admitting a maximal $G_2$ acting on spacelike surfaces. 
The whole family is self-similar in the sense that it has a further
homothetic vector field and it contains subfamilies of both tilted and 
non-tilted Bianchi models. This is the first inhomogeneous family with a
$\gamma$-law equation of state having a free $\gamma$ which includes
the $\gamma=0$ case, something which may be very useful in order to study
perturbations of the dust case.

The structure of the paper is as follows. In Section \ref{sec:models}
we introduce the line-element for the whole family
in non-comoving canonical coordinates and show
the ranges of the free parameters and the perfect-fluid variables,
which are in turn expressed in terms of its velocity potential
(subsections \ref{sec:lineel} and \ref{sec:pf}).
The kinematical quantities
of the fluid flow and their properties, as well as the deceleration 
parameter and the Weyl tensor are given in subsections
\ref{sec:solgammaro} and \ref{sec:weyl}.
Then, in subsection \ref{sec:part}, we study the general symmetries of 
the spacetimes including the analysis of the special cases
that arise, which include previously known exact `non-tilted'
Bianchi spacetimes together with some other {\it new} `tilted' ones.
Next, comoving canonical coordinates are introduced in subsection
\ref{sec:comoving}, on the one hand to show that the metric is non-separable
in comoving coordinates in general, and on the other hand to
construct a half-null coordinate
system that will be used to make manifest the singularity structure and
its type
in subsection \ref{sec:sing}.
The result is that the whole class
presents a universal spacelike big-bang singularity
in the finite past, which turns out to be of Kasner type \cite{lifkala,liang}.
Similarly, the future asymptotic behaviour
of the solutions is shown in subsection \ref{sec:future}.

In Section \ref{sec:subfam} we present the most interesting
particular subfamilies and limits of the general spacetime.
In particular, two vacuum limits in the half-null coordinates
are found in subsection \ref{sec:limits}, providing two two-parameter
families of pure gravitational pp-wave solutions.
The $p=\varrho /3$ subfamily is studied then in subsection \ref{sec:radia},
and it is given the interpretation of a Bianchi V background inhomogenized
by means of a pure gravitational wave
travelling along the direction orthogonal
to the surfaces of transitivity of the $G_{2}$ group. This 
gravitational wave is closely related to some acoustic waves which
travel along and tilt the perfect fluid. This result is new in
the sense that all previous works concerning propagation of 
waves in curved backgrounds were developed in the case of vacuum or 
massless minimally coupled scalar fields without potential
(the latter includes the stiff fluid $p=\varrho$ case,
see \cite{wainmarsh} for a pioneering treatment of the subject), see e.g.
\cite{CCM,Adams1,verdaguer,bicakgrif} and
references therein for a selection of main results,
or in an anisotropic generalization of the stiff fluid
in which the energy density equals
the pressure on the direction of propagation
of the waves \cite{letelier1}, 
and finally, in the case of electromagnetic fields, see \cite{alekseev}.
There have also been other
works on solutions describing exact solitonic perturbations of
$\gamma$-law perfect fluid backgrounds
\cite{ibver,diglepul,verdaguer}, but the formalism
consists in translating the solutions
of the Einstein field equations into equivalent five-dimensional massless
scalar field spacetimes, and thus the backgrounds are severely
restricted by some
conditions on the matter content so that eventually only
Friedmann-Lema\^{\i}tre-Robertson-Walker (FLRW)
spacetimes were used.
Moreover, these perturbations give rise
to anisotropies in the energy-momentum tensor.
In our case, the
gravitational wave is exact and travels on a spatially homogeneous but
anisotropic background, and both the background and the resulting inhomogeneous
spacetime satisfy the same realistic equation of state $p=\varrho /3$.
This is the first known example of such a situation.

Finally, the dust 
subfamily is identified within the general classes found in 
\cite{SEVEdust} in subsection \ref{sec:dust}.
Throughout the paper we follow the following conventions and
notations.
The metric has signature $+2$. $\lie_{\vec v}$ denotes
the Lie derivative with respect to the vector field $\vec v$.
Primes and dots will stand for derivatives with respect to $x$ and $t$
respectively. Greek indices run from 0 to 3.
We take units with $8\pi G =c =1$.

\section{The models}
\label{sec:models}
This section is devoted to presenting, in as compact a way as 
possible, the new family of spacetimes and their main geometrical and 
physical properties. 

\subsection{The line-element}
\label{sec:lineel}
The line-element was derived using separability
of the metric functions in non-comoving coordinates \cite{G2NC}
(more precisely, it arises as a particular case of the $m=n=2$ (22BIIc) 
class as defined in \cite{Raultesi}) and it 
can be written as follows
\bea
&&\d s^2=F^2(t,x)\left(-\d t^2+\d x^2\right)+e^{\mu \frac{a}{a+b}(t-x)}
\fume \nonumber\\
&&\hspace{1cm}
\times\left[\left(e^{\mu \frac{b}{a+b}(t-x)}\fume\right)^{2l}
\d y^2+\left(e^{\mu \frac{b}{a+b}(t-x)}\fume\right)^{-2l}\d z^2\right],
\hspace{1cm}
\label{eq:ds2gamma}
\eea
with
\[
F(t,x)\equiv \exp\left[\frac{\mu}{a+b}(td-cx)\right]\fume^{\lambda},
\]
and where we have defined the following constants:
\bean
&&a\equiv (4\lambda+1)(\nu^2-1)(\nu-1),\\
&&b\equiv (6\lambda\nu-2\lambda+\nu+1)(\nu^2-1),\\
&&c\equiv \nu(2\lambda+1)(\nu^2+1+2(2\nu-1)(\nu+1)\lambda),\\
&&d\equiv \nu^2+1+2(\nu+3)\lambda\nu^2+2(5\nu^2-4\nu+1)(\nu+1)\lambda^2,
\eean
with $\lambda$ being
$$\lambda\equiv l^2-1/4$$
an auxiliary constant that will be used for the sake of simplicity.
Actually, the constants satisfy the relation
$c=\nu(d-\lambda(a+b))\equiv \nu \hat{c}$
that allows to cast the function $F(t,x)$ in
the alternative and possibly more convenient form given by
$$
F(t,x)=\exp\left[\frac{\mu\hat{c}}{a+b}(t-\nu x)\right]
\left(e^{\mu t}-1\right)^\lambda.
$$
Nevertheless, and for the sake of simplicity in some expressions,
we prefer to keep the four constants $a,b,c,d$ and $F$ as given previously.

The family of solutions has then three free parameters, $l$ (or $\lambda$),
$\nu$ and $\mu$, although the latter simply provides the coordinate
scaling.
We obviously have $\lambda\geq -1/4$, and we can choose $\mu>0$ 
without loss of generality (see below). Furthermore, we must have $a+b\neq 0$,
which eventually will be equivalent to
$$
(5\nu-3)\lambda+\nu\neq 0 .
$$

\subsection{The perfect fluid}
\label{sec:pf}
The line-element (\ref{eq:ds2gamma}) is a solution of Einstein's field 
equations for a perfect-fluid energy-momentum tensor $T^{}_{\alpha\beta}=
(\varrho+p)u^{}_\alpha u^{}_\beta+
p \g^{}_{\alpha\beta}$ ($\varrho+p\neq 0$) whenever $\nu$ is restricted by 
\be
1-\nu^2>0,
\label{eq:nu}
\ee
which in turn implies the last condition of the previous subsection.
The unit velocity vector field $\vec u$ reads then
\[
\vec u=\frac{1}{F\sqrt{1-\nu^2}}\left[ \Part +\nu\Parx \right].
\] 
The ranges for $\nu$ and $\lambda$ immediately imply $a>0$. One can 
also deduce that $d>0$ as follows: we have 
$5\nu^2-4\nu+1>0$, $\forall \nu$, and thus
$d>d |_{\lambda=-1/4}=(3-\nu)(3-\nu^2)/8>0$.

The energy density is given by
\[
\varrho=F^{-2}\left(\frac{a-b}{a+b}\right)\frac{\mu^2(\lambda+1)}{\fuma},
\]
and the equation of state is barotropic and obeys the gamma law
\[
p=\gamma \varrho,
\]
where $\gamma$ is given explicitly in terms of $\lambda$ by
\[
\gamma=\frac{\lambda}{\lambda+1}=\frac{l^2-1/4}{l^2+3/4},
\]
so that we have $-1/3\leq \gamma<1$.
Notice that in the $\varrho+3p=0$ case
$\lambda=-1/4 \Leftrightarrow l=0$ and the solutions
admit a plane $G_3$ on $S_2$.

 From the above expressions is clear that the
solutions have an initial big-bang singularity at $t=0$,
and this is why we have taken $\mu>0$ without loss of generality.
Subsection \ref{sec:sing} is devoted to studying
the singularity structure of the family, and in particular it will be 
shown that the big-bang singularity at $t=0$ is the only one
for the whole family.

The perfect-fluid region covers the entire manifold,
and we have $\varrho>0$ everywhere whenever $a^2-b^2>0$,
which is equivalent to
\be
(5\nu-3)\lambda+\nu <0,
\label{eq:positi}
\ee
where we have taken into account that $(a-b)/(2(1-\nu^2))=(\nu+1)\lambda+1>0$
(so that $\varrho\neq 0$) which follows from the ranges for $\nu$ and
$\lambda$.
In fact, the previous condition (\ref{eq:positi}) implies
$a+b=-2(1-\nu^2)\left[(5\nu-3)\lambda+\nu\right]>0$, and this
ensures the fulfillment of both the
dominant and the strong energy conditions on the whole spacetime.

Since the fluid flow is irrotational it can be expressed
as the normalized gradient of a scalar field, the so-called
velocity potential $\sigma$ \cite{liang}, that is
$$
u_{\alpha}=\sigma_{,\alpha}/\sqrt{-\sigma_{,\beta}\sigma^{,\beta}},
$$
where the commas stand for the partial derivative.
Because of the gamma law equation of state, the
corresponding energy density reads
$\varrho=(-\sigma_{,\alpha}\sigma^{,\alpha})^{(\gamma+1)/2\gamma}$,
and the velocity potential satisfies a homogeneous wave equation
(non-linear whenever $p\neq \varrho$) which gives the sound wave
equation once it is linearized \cite{liang}.
The velocity potential for the whole family is given by
$$
\sigma=-
\left(\mu^2(\lambda+1)\frac{a-b}{a+b}\right)^{\frac{\lambda}{2\lambda+1}}
\frac{(2\lambda+1)(a+b)}{\mu \hat{c}\sqrt{1-\nu^2}}
\exp\left[\frac{\mu \hat{c}}{(2\lambda+1)(a+b)}(t-\nu x)\right]
$$
apart from an additive constant,
so that the parameter $\nu$
is nothing but the peculiar spatial fluid velocity $-\sigma'/\dot{\sigma}$.
The range given in (\ref{eq:nu}) for the perfect fluid is consistent with
this interpretation, so that given any $\gamma$
the corresponding subfamily contains all the possible values the ratio
$\sigma'/\dot{\sigma}$ can achieve.
Indeed, since this ratio is constant,
this family of solutions does not follow an asymptotically
velocity-dominated regime near the initial
singularity except for the cases $\nu\rightarrow 0$,
which could be seen as perturbations of homogeneous models
(see section \ref{sec:part}).

\subsection{The kinematical quantities}
\label{sec:solgammaro}
In order to compute the kinematical (and other) quantities for the 
fluid congruence defined by $\vec u$, let us take the orthonormal tetrad
$\{\bm{\theta}^\alpha\}$ with 
$\bm{\theta}^\alpha \propto \d x^\alpha$ in the above coordinate 
system $\{x^{\alpha}\}=\{t,x,y,z\}$. Of course, the vorticity of the 
fluid congruence vanishes identically. Regarding its acceleration, its
non-vanishing components are
\be
\acc_0=-\nu\acc_1,\hspace{1cm}
\acc_1=F^{-1} \mu\lambda \frac{\nu}{1-\nu^2}\;
\frac{1}{\fume},
\label{eq:pgammaacc}
\ee
so that the fluid flow does not follow geodesic trajectories
at any point of the spacetime, except for the special cases
$\nu=0$ or $\lambda=0$, in which the acceleration vanishes everywhere.
These special cases will be discussed later in subsection \ref{sec:part}.

With respect to the expansion and the non-zero components of the
shear tensor we have
\be
\theta=F^{-1}\frac{\mu}{\sqrt{1-\nu^2}(a+b)\fuma}
\left[\alpha^2\fuma+(a+b)(\lambda+1)\right],
\label{eq:expansion}
\ee
\bean
&&\sigma_{00}=\nu^2\sigma_{11},\hspace{1cm}
\sigma_{01}=-\nu\sigma_{11},\\
&&\sigma_{11}=-F^{-1}
\frac{\mu\left[a\nu+b-e^{\mu t}a(\nu-1)\right]}
{(1-\nu^2)^{3/2}(a+b)\fuma}+\frac{2}{3(1-\nu^2)}\theta,\\
&&\sigma_{22}+\sigma_{33}=F^{-1}\frac{\mu}{3\sqrt{1-\nu^2}(a+b)\fuma}\\
&&
\hspace{2cm}\times
\left\{[-3(d-\nu c)-\alpha^2]\fuma-(a+b)(2\lambda-1)\right\},\\
&&\sigma_{22}-\sigma_{33}=F^{-1}\frac{2\mu l
\left[a+\nu b-e^{\mu t}b(\nu-1)\right]}{\sqrt{1-\nu^2}(a+b)\fuma},
\eean
where we have defined $\alpha^2\equiv d-\nu c + a(1-\nu)$, which is indeed
a positive constant for the given ranges of $\nu$ and $\lambda$.
This can be easily deduced from its explicit expression
\[
\alpha^2 = 2(1-\nu^2)(1+\lambda)\left[(4\nu^2-3\nu+1)\lambda+(\nu^2)+(1-\nu)
\right],
\]
as every term between round brackets is strictly positive because of
(\ref{eq:nu}), (in fact $4\nu^2-3\nu+1>0$, $\forall \nu$), and so
the less favorable case would correspond to $\lambda=-1/4$, which gives a
positive value for the term in square
brackets for the valid range of $\nu$.

Therefore, from expression (\ref{eq:expansion})
we see that the fluid congruence is expanding everywhere,
$\theta >0$, starting with an unbounded value at the initial singularity
$t=0$ and decreasing  continuously from then on arriving eventually
to zero as $t$ tends to infinity.
A straightforward calculation shows also that
\bean
&&\frac{\sigma_{\alpha\beta}\sigma^{\alpha\beta}}{2\theta^2}
(t\rightarrow \infty)\longrightarrow
\frac{2(1-\nu^2)^2}{3\alpha^2}
\left[
\frac{1}{4}\left( a(1-\nu)-2(d-\nu c)\right)^2+3(1-\nu)^2 b^2 l^2
\right],
\eean
so that the only case in which the solutions isotropize in the
future is given by $\lambda=1/2$ and $\nu=0$ ($\sigma_{11}=0$,
$\sigma_{22}=-\sigma_{33}\neq 0$), which is a `comoving' family with
an additional symmetry, as we will see later in subsection 
\ref{sec:part}.

Finally, we present the expression of the deceleration parameter $q$,
whose general definition is
\[
\vec u (\theta^{-1})\equiv \frac{1}{3}(1+q),
\]
so that it reads
\[
\frac{1}{3}(1+q)=\frac{\mu}{F\sqrt{1-\nu^2}\theta}
\left[\frac{d-\nu c}{a+b}+\frac{e^{\mu t}+\lambda}{\fuma}
-\frac{\alpha^2 e^{\mu t}}{\alpha^2\fuma+(a+b)(\lambda+1)}
\right],
\]
from where it can be checked that at the singularity $q(t\rightarrow 0)=2$. It 
is interesting to remark that, as follows from the previous expression 
and (\ref{eq:expansion}), $q$ is independent of $x$, despite the 
inhomogeneity of the solutions. It should be stressed that this 
result holds in the above coordinate system, which is {\it not} 
comoving, and it is a simple consequence of the separability of the 
metric components in these coordinates.
In the comoving coordinates adapted to the fluid flow, 
which will be given in subsection \ref{sec:comoving}, the deceleration 
parameter certainly depends on the corresponding spatial variable.

\subsection{The Weyl tensor}
\label{sec:weyl}
Concerning the Weyl tensor, the non-vanishing scalars
computed in the null tetrad (see \cite{KRAM})
$\bm{k}=2^{-1/2}(\bm{\theta}^0-\bm{\theta}^1)$,
$\bm{l}=2^{-1/2}(\bm{\theta}^0+\bm{\theta}^1)$,
$\bm{m}=2^{-1/2}(\bm{\theta}^2+i \bm{\theta}^3)$,
are given by
\bean
&&\Psi_0-\Psi_4=- \frac{\mu^2 l}{F^2(a+b)^2\fuma}\\
&&\hspace{2cm}\times
\left[\left(2(b\lambda+c)-a\right)a+\left((2\lambda-1)b-2d\right)b+
2 e^{\mu t}(c+d-a)b\right],\\
&&\Psi_2=\frac{\mu^2}{12(a+b)F^2\fuma^2}
\left[e^{\mu t}\left((10\lambda+1)b+(2\lambda-1)a\right)+
(a-b)(4\lambda+1)\right],\\
&&\Psi_0=\frac{\mu^2 l}{2(a+b)F^2\fuma^2}
\left[(a+b)(2\lambda-1)+2(c-d)-e^{\mu t}\left(2(c-d)-b-a\right)\right],
\eean
and therefore, the Weyl tensor does not vanish
and there is no possible particularization
to a conformally flat solution within the whole family.
In particular, flat spacetime is not included in 
the family.

The Petrov type is I at generic points for the general case with a 
maximal $G_2$, and type D for the $G_{3}$ on $S_{2}$
case ($l=0$) as well as for the case $\nu=0$, $\lambda=2$,
which admits two additional isometries,
belonging then to the class of LRS models (see next subsection
\ref{sec:part}).

The non-zero component of the magnetic part of the Weyl
tensor with respect to $\vec u$ is given by
$$
H^{}_{23}(\vec u)=\frac{1+\nu}{1-\nu}\;\Psi_0-\frac{1-\nu}{1+\nu}\;\Psi_4,
$$
so that $H(\vec u)$ only vanishes in the type D cases: $\nu=0,\lambda=2$
or $l=0$.

\subsection{The symmetries. Special cases}
\label{sec:part}
The line-element (\ref{eq:ds2gamma}) admits in general a 2-parameter 
group of isometries, generated by the two Killing vectors,
$$
\vec{\xi}=\frac{\partial}{\partial y}\, , \hspace{1cm}
\vec{\eta}=\frac{\partial}{\partial z}\, ,
$$
which obviously commute, so that the $G_{2}$ is Abelian. Moreover, 
the metric admits a homothetic vector field given by
\be
\vec{\zeta}=2\Parx+\frac{\mu}{a+b}(a+2bl-2c)y\Pary
+\frac{\mu}{a+b}(a-2bl-2c)z\Parz,
\label{eq:kill}
\ee
which satisfies
$$\lie_{\vec{\zeta}}\; \g^{}_{\alpha\beta}=
-4c\mu \g^{}_{\alpha\beta}.$$
It follows that the general family of solutions with $c\neq 0$ belongs to
the class of so-called `tilted' inhomogeneous self-similar 
perfect-fluid models \cite{ali}:
the velocity vector $\vec u$ is neither tangential nor orthogonal to the orbits
of the 3-dimensional homothetic
group $H_3$ generated by $\{\vec \xi,\vec \eta,\vec \zeta\}$,
which is acting on spacelike hypersurfaces ($S_3$).
The algebraic structure of $H_3$
is described by the following Bianchi types:
Bianchi VI$_h$ with $h=-\left[(a-2c)/2lb\right]^2$
(Bianchi III is indeed included when $h=-1$)
whenever $lb\neq 0$, and Bianchi V when $lb=0$. In this last 
possibility, by taking into account that the case $l=0$
already admits a $G_3$ on $S_2$ group of isometries,
the resulting group is an $H_4$ acting on $S_3$ if $l=0$.

Coming to the possible particular cases with further symmetry we
first find the already mentioned case with $l=0$ ($\lambda=-1/4$),
so that the equation of state has the
form $\varrho+3p=0$. This solution admits a $G_3$ group of isometries
acting multiply-transitively on spacelike plane $S_2$-orbits, and
the Petrov type is D.

When $l\neq0$, the only possible cases with additional Killing vectors are 
all given by $c=0$, in which case the line-element (\ref{eq:ds2gamma})
admits the vector field $\vec{\zeta}$ given by (\ref{eq:kill})
restricted to $c=0$, that is
\[
\vec{\zeta}=2\Parx+\frac{\mu}{a+b}(a+2bl)y\Pary +\frac{\mu}{a+b}(a-2bl)z\Parz,
\]
so that the $H_3$ becomes a $G_3$ on $S_3$ with the same Bianchi types
as indicated above (see \cite{EM,ryan,Llibre} and references therein).
The explicit expression of $c$ leads to two possibilities for $c=0$:
\begin{enumerate}
\item Case with $\nu=0$. Now the velocity vector $\vec u$ is orthogonal to
the orbits of the simply-transitive $G_3$
group, so that the resulting solutions belong to the following
{\em non-tilted\/} Bianchi classes of spacetimes:
\begin{itemize}
\item Bianchi V when $\lambda=1/2$. This homogeneous spacetime is a special
case
of the general Bianchi V family with $p=\varrho/3$
found by Ruban in \cite{ruban}
(line-element (9.20) in \cite{Llibre} with $\alpha=m^2$).
This is the only solution of the whole family such that the flow generated
by $\vec u$ isotropizes in the future (see subsection \ref{sec:solgammaro}).
\item Bianchi III when $\lambda=2$. As already mentioned,
this case corresponds to another
algebraically special solution (Petrov type \D),
which actually admits a fourth Killing vector
given by
\[
\vec{\chi}=y\left(\Parx+\frac{3}{8}\mu\Pary\right),
\]
so that this case belongs to the LRS ($G_4$ on $S_3$) models
of class II in \cite{SteEll}. As we said in subsection 
\ref{sec:solgammaro}, this is the only solution in
(\ref{eq:ds2gamma}), apart from the plane $G_3$ on $S_2$ case
($l=0$), having a purely electric Weyl tensor with respect 
to the fluid vector $\vec u$, $H(\vec u)=0$ (see \cite{henkEll}).
\item
A one-parameter family of Bianchi VI$_h$ spacetimes when $\lambda\neq 1/2,2$.
This family is included in the class of evolving non-tilted Bianchi VI$_h$
spacetimes for $p=\gamma\varrho$ (table 9.4 in \cite{Llibre}).
Following the notation in \cite{Llibre} (using a tilde for the 
quantities in \cite{Llibre}),
the present families are included in the
general cases with $\tilde{k}=|4-3\tilde{\gamma}|/2$
(table 9.4 in \cite{Llibre}), discovered by Uggla and Rosquist \cite{UgRo}
up to quadratures.
\end{itemize}
\item Case with
\[
\lambda=\frac{1+\nu^2}{2(1-2\nu)(\nu+1)},
\]
which satisfies $\varrho>0$ and $1/\sqrt{10}<\gamma<1$
(for $\nu\in (-1,1/2)$).
In this case, and for $\nu\neq 0$, the perfect-fluid flow
has a non-vanishing projection onto the $G_3$ orbits,
so that it constitutes a
one-parameter family of exact `tilted' Bianchi solutions.
The free parameter can be chosen to be $\gamma$ with the restriction
above and taking into account that the case $\gamma=1/3$ (and $\nu=0$)
falls onto the previous `non-tilted' Bianchi V case.
The Bianchi type for this `tilted' homogeneous solutions is VI$_h$
with $h=-[\nu^2(4\lambda+1)]^{-1}$ (including Bianchi III).
\end{enumerate}

\subsection{The comoving coordinates}
\label{sec:comoving}
As is known \cite{WW1}, every perfect-fluid diagonal $G_2$ solution
can be written in comoving coordinates $\{T,X,y,z\}$ keeping the
diagonal form of the metric. By comoving coordinates we mean those 
such that $\vec{u}\propto \partial_{T}$.
The comoving coordinates will be useful for the study of the singularities
that will be performed in subsection \ref{sec:sing}. Also, this will 
prove that the use of comoving coordinates may sometimes be not well 
adapted to writing some solutions in explicit form, or even to look 
for them.

The explicit change to comoving coordinates for the metric
(\ref{eq:ds2gamma}) is easily found to be
\be
t=\frac{1}{\sqrt{1-\nu^2}}\left(T +\nu X\right),
\hspace{1cm}
x=\frac{1}{\sqrt{1-\nu^2}}\left(X +\nu T\right), \label{comov}
\ee
where the Jacobian of the change is 1.
By writing $\hat{\mu}\equiv \mu/\sqrt{1-\nu^2}$, the line-element
becomes
\bea
&&\d s^2=F^2(T,X)\left( -\d T^2+\d X^2\right)+
e^{\hat{\mu} \frac{a}{a+b} (1-\nu)(T-X)}J(T,X)
\nonumber\\
&&\hspace{1cm}\times
\left( \left(e^{\hat{\mu} \frac{b}{a+b} (1-\nu) (T-X)}J(T,X)\right)^{2l}\d y^2+
\left(e^{\hat{\mu} \frac{b}{a+b} (1-\nu) (T-X)}J(T,X)\right)^{-2l}\d z^2\right)
\label{eq:ds2comov}
\eea
where now
\[
F(T,X)=\exp\left[\frac{\hat{\mu}}{a+b}
\left((d-c\nu)T-(c-\nu d)X\right)\right]J^\lambda(T,X),
\]
and we have that
\be
J(T,X)=1-\exp\left[-\hat{\mu}(T+\nu X)\right].
\label{eq:J}
\ee
Then the fluid velocity vector field simply reads $\vec u=F^{-1}\partial_{T}$.
As is obvious, this family of solutions is not separable in comoving
coordinates for $\nu\neq 0$.
See \cite{Raultesi} for a study of the loss
of separability when performing arbitrary coordinate changes (within the
2-spaces orthogonal to the $G_{2}$-group orbits) which keep
the diagonal form of the metric.

As we can see, the structure of the line-element is perhaps not
too cumbersome in comoving coordinates, but it is complicated enough 
so that the solutions had not been found until the Ansatz of separability
in non-comoving coordinates was used.
The structure shown in (\ref{eq:ds2comov}) may indicate
some new Ansatzs providing, perhaps,
generalizations of these $G_2$ spacetimes with $p=\gamma \varrho$.

\subsection{The half-null coordinates. Singularity structure}
\label{sec:sing}
As mentioned previously, the solutions present
an initial big-bang singularity at the spacelike hypersurface $t=0$
coming from the vanishing
of the function $J$ given in (\ref{eq:J}) for the metric
(\ref{eq:ds2comov}). Nevertheless,
the form of the function $F/J^\lambda$ may suggest
that other singularities could be present, as for instance at
$x\rightarrow \pm \infty$.
We are going to show that this is not the case,
and consequently the only singularity of the solutions
is the reachable universal spacelike singularity at $t=0$,
which in the comoving coordinates is given by $T+\nu X=0$.

To that end, let us start by noticing that the coordinate ranges of
(\ref{eq:ds2comov}) are in principle only restricted by
\be
T+\nu X>0.
\label{eq:range}
\ee
Simple inspection
on the expressions for the energy density and the Weyl scalars
indicates that in this coordinate range the only other
possible singular points
would be those where the function $F/J^\lambda$ vanishes,
and that the singularities of both the Ricci and Weyl
tensors coincide.
At this point, it is very useful to perform the change to null
coordinates $\{U,V\}$ in the surfaces orthogonal to the orbits
of the $G_{2}$ group, given by
\be
U=\frac{1}{\sqrt{2}}\left(T-X\right),\hspace{1cm}
V=\frac{1}{\sqrt{2}}\left(T+X\right),
\label{eq:UV}
\ee
so that (\ref{eq:range}) now becomes
\be
V(1+\nu)+U(1-\nu)>0.
\label{eq:range2}
\ee

Let us define now the following two constants
$$
\kappa_u\equiv \sqrt{2}\frac{\hat{\mu}}{a+b}(1-\nu)(d+c),\hspace{1cm}
\kappa_v\equiv \sqrt{2}\frac{\hat{\mu}}{a+b}(1+\nu)(d-c),
$$
whose fundamental property will turn out to be their
positivity. Indeed, we have first of all that
$$
(1-\nu)(d+c)=(1-\nu^2)\left[2(3\nu-1)^2\lambda^2+(8\lambda+1)\nu^2+1\right].
$$
Since $\lambda\geq -1/4$ and $\nu^2<1$ we have that $\nu^2(8\lambda+1)>-1$,
which, together with $a+b>0$, easily leads to $\kappa_u>0$.
On the other hand, the following explicit expression
$$
(1+\nu)(d-c)=(1-\nu^2)\left[2\lambda^2(1-\nu^2)+(4\lambda+1)\nu^2+1\right],
$$
directly shows that $\kappa_v>0$ too.
The function $F$ is now such that
$$
F(U,V)J^{-\lambda}(U,V)=
e^{\frac{1}{2}\left(\kappa_u U+\kappa_v V\right)},
$$
and the usefulness of the change is now clear
since the region with $F/J^\lambda\rightarrow 0$,
that is $\kappa_u U+\kappa_v V\rightarrow -\infty$,
can be reached within the range (\ref{eq:range2})
only if $U\rightarrow -\infty$ or $V\rightarrow -\infty$.
In order to ascertain whether or not they are reachable,
let us then bring them to finite values by making
the typical coordinate change
$$
U=\frac{1}{\kappa_u}\log\ku,\hspace{1cm}
V=\frac{1}{\kappa_v}\log\kv,
$$
so that the range of the new null coordinates $(u,v)$ is given, in
principle, by $u>0$, $v>0$ and the restriction coming from (\ref{eq:range2}),
which reads
\be
\ku^{\frac{1-\nu}{\kappa_u}}\kv^{\frac{1+\nu}{\kappa_v}}
>1.
\label{eq:range3}
\ee
The line-element becomes then
\bea
&&\d s^2=-2J^{2\lambda}(u,v)\d u \d v+\ku_{}^\frac{a}{d+c}J(u,v)
\nonumber\\
&&\hspace{2cm}
\times\left\{\left[\ku_{}^\frac{b}{d+c}J(u,v)\right]^{2l} \d y^2+
\left[\ku_{}^\frac{b}{d+c}J(u,v)\right]^{-2l} \d z^2\right\},
\hspace{1cm}
\label{eq:ds2hn}
\eea
where now
$$
J(u,v)=1-\ku^{-\frac{a+b}{2(d+c)}}
\kv^{-\frac{a+b}{2(d-c)}},
$$
and $F/J^\lambda=\ku \kv$, so that the other possible singularities
have been transported to $uv=0$ in the new coordinates.

\begin{figure}[p]
\centering
\begin{picture}(60,48)
\put(0,40){$u$}
\put(60,35){$v$}
\put(31,48){$\hat{t}$}
\put(57,4){$\hat{x}$}
\mbox{\epsfig{file=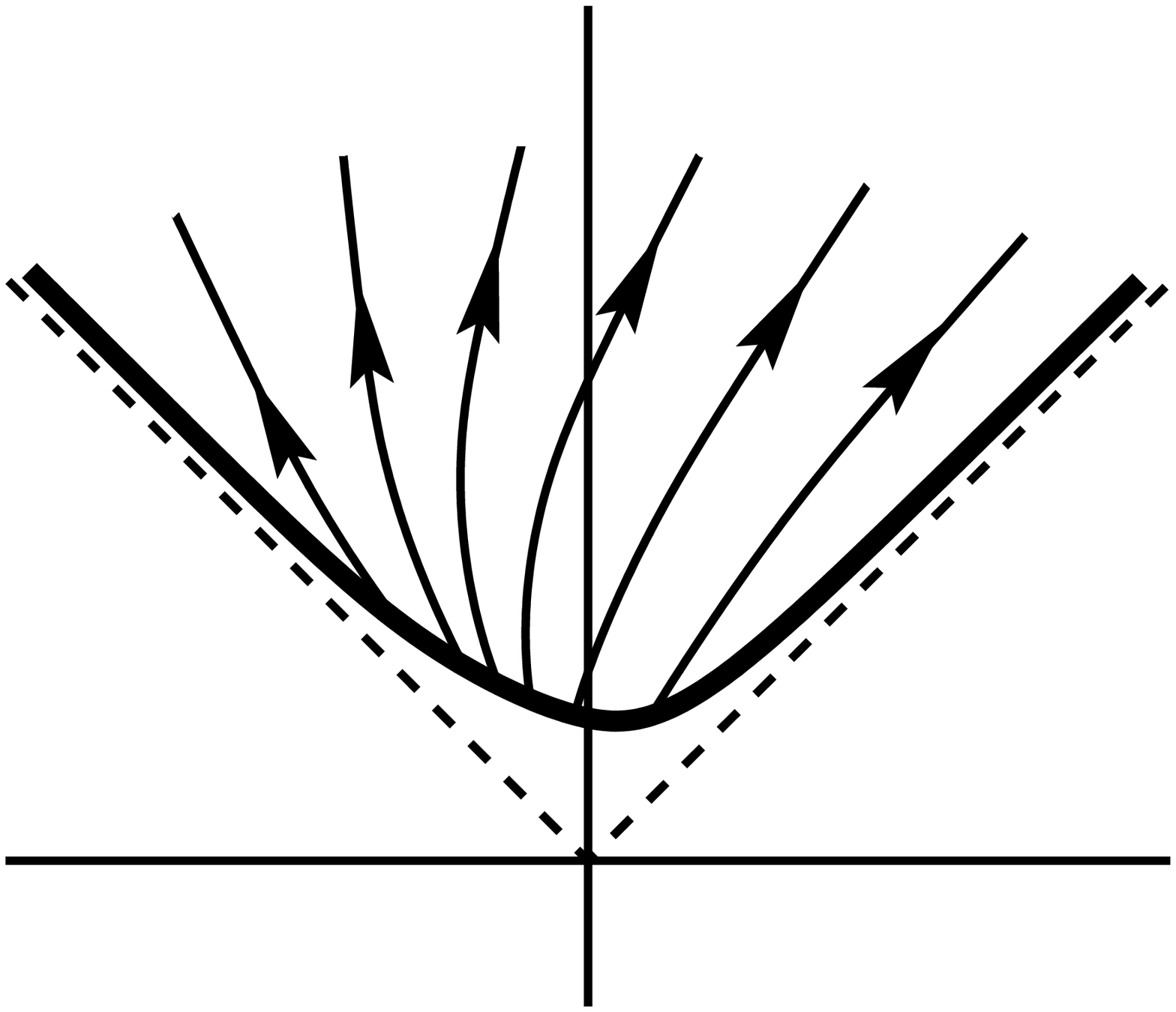,width=6cm}}
\end{picture}
\hspace{1cm}
\begin{picture}(60,48)
\put(0,40){$u$}
\put(60,35){$v$}
\put(31,48){$\hat{t}$}
\put(57,4){$\hat{x}$}
\mbox{\epsfig{file=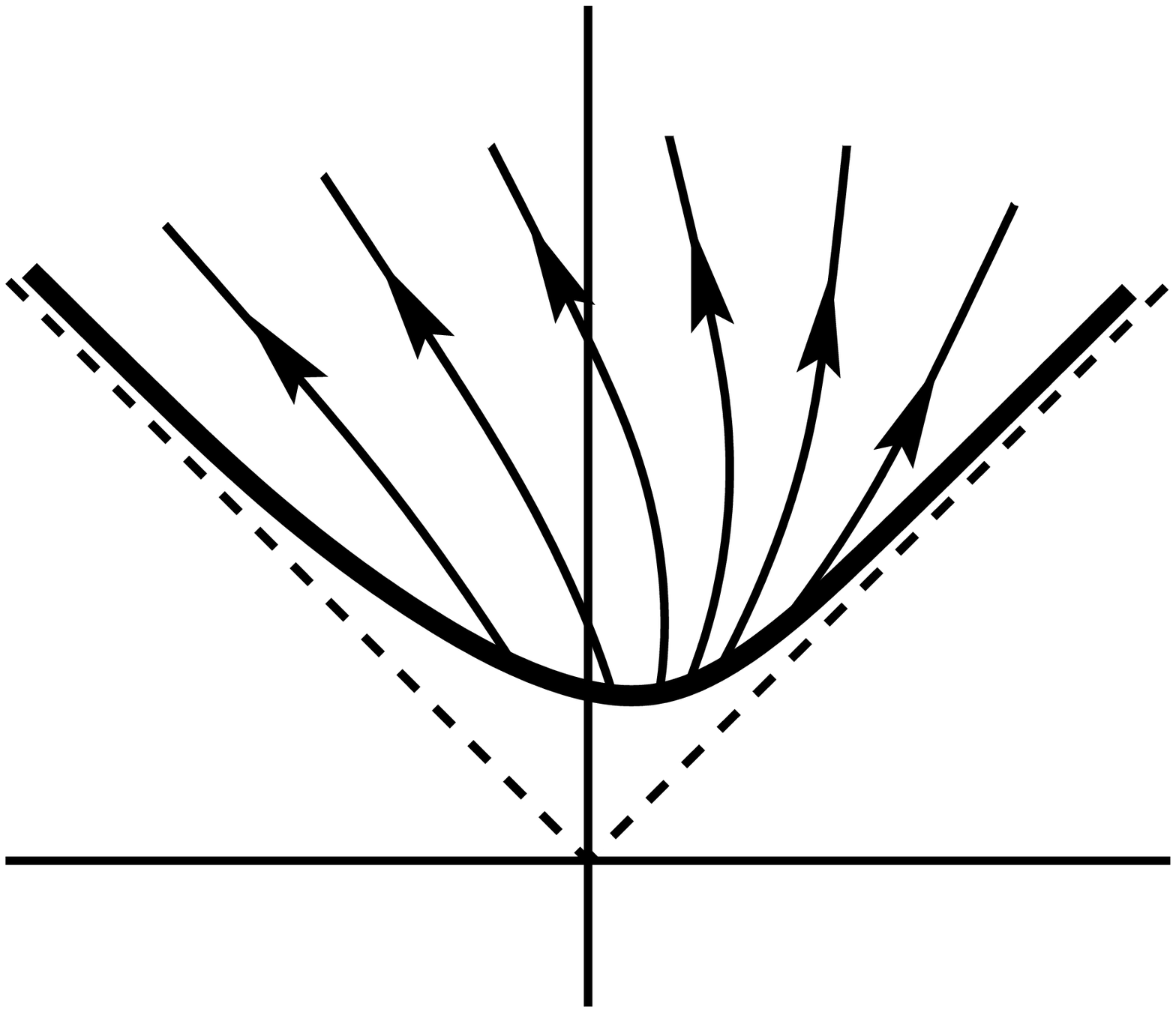,width=6cm}}
\end{picture}
\centering\parbox{6cm}{\centering(a)}\hspace{1cm}\parbox{6cm}{\centering(b)}\\
\centering
\begin{picture}(60,48)
\put(0,38){$u$}
\put(60,33){$v$}
\put(31,48){$\hat{t}$}
\put(57,3){$\hat{x}$}
\mbox{\epsfig{file=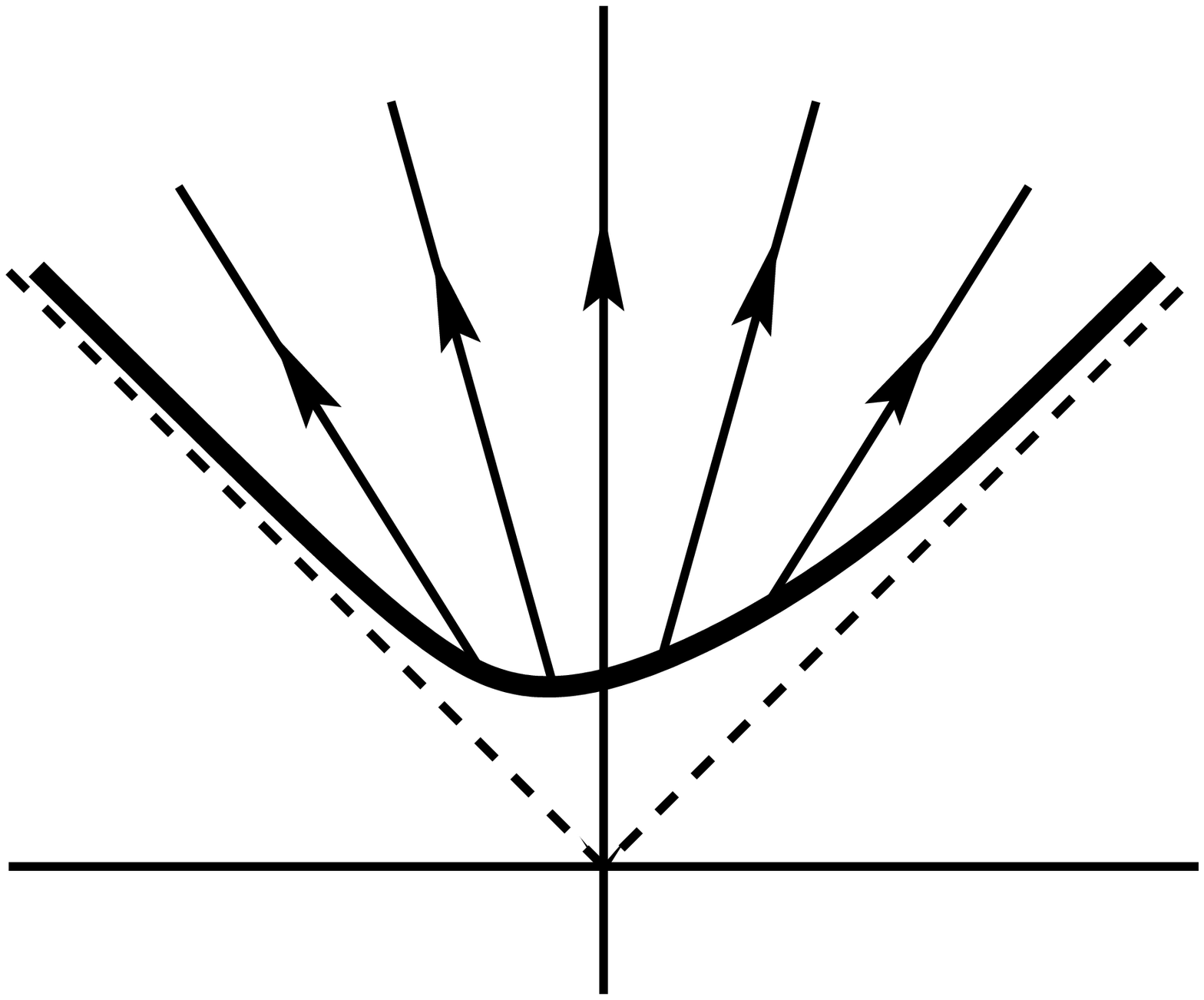,width=6cm}}
\end{picture}\\
\centering\mbox{\hspace{6cm}}\mbox{(c)}\mbox{\hspace{6cm}}
\caption{Diagrams showing the singularity in the $(u,v)$-surfaces
for the three possibilities (a) $\kappa_v>\kappa_u$,
(b) $\kappa_v<\kappa_u$, (c) $\kappa_v=\kappa_u$.
The whole spacetime is the product of these surfaces with the group orbits.
As usual, null lines are at 45$^o$.
The fluid flow is indicated by arrowed lines in the region given by
expression (\ref{eq:range3}). Notice that
the shown coordinates $\{\hat{t},\hat{x}\}$ would correspond
to the Lorentzian coordinates related to $\{u,v\}$ in the same
manner as $\{T,X\}$ are related to $\{U,V\}$ (\ref{eq:UV}).                   
The big-bang singularity at $t=0$, or equivalently (\ref{eq:singuv}), which
is a spacelike hypersurface, is denoted
by a thick curve which tends asymptotically to
$uv=0$ (denoted by dashed lines).
Of course, it is evident that $uv=0$ is hidden `in the past'
of the big-bang singularity in the physically meaningful spacetime
($\varrho>0$), and therefore the values $u=0$ or $v=0$ are unreachable.
}
\label{fig:main}
\end{figure}

The point now is that the singularity at $t=0$,
which lies in the limit of the restriction given in (\ref{eq:range3})
and has the form (for some constant $A$)
\be
u=A^2 v^{-(d+c)/(d-c)},
\label{eq:singuv}
\ee
{\it does hide} the other possible singularities at $uv=0$
in the sense that any endless past-directed causal curve from any point
in our manifold terminates necessarily at $t=0$
(this is why it is called a universal big-bang singularity, 
see \cite{totxosing}). In other words, $uv=0$ is not accessible within
the physical spacetime, see figure \ref{fig:main}.

\subsubsection{Singularity type}
The three (non-vanishing)
eigenvalues of the distortion tensor of the fluid congruence
defined by $\vec u$,
$$
\theta_{\alpha\beta}\equiv
\sigma_{\alpha\beta}+\frac{1}{3}\theta P_{\alpha\beta},
$$
being
$P_{\alpha\beta}\equiv g_{\alpha\beta}+u_\alpha u_\beta$ the
projector orthogonal to $\vec u$, read as follows at the limit
$t\rightarrow 0$:
$$
\theta_1=f\;\lambda,\hspace{1cm}
\theta_2=f\;\left(\frac{1}{2}+l\right),\hspace{1cm}
\theta_3=f\;\left(\frac{1}{2}-l\right),
$$
where we have defined $f\equiv \lim_{t\rightarrow 0} \mu/(\sqrt{1-\nu^2}F J)$.
The behaviour of the fluid congruence is described then by the
scale factors $l_i$ $(i:1,2,3)$ defined by $\vec u(\log l_i)=\theta_i$,
giving three different possibilities depending on the values of $\lambda$
at the limit $t\rightarrow 0$:
\begin{itemize}
\item $\lambda<0$ : $l_1\rightarrow \infty$, $l_2,l_3\rightarrow 0$.
\item $\lambda=0$ : $l_2\rightarrow 0$, $l_1$ and $l_3$ tend to a finite
value.
\item $\lambda>0$ : $l_3\rightarrow \infty$, $l_1,l_2\rightarrow 0$.
\end{itemize}
Therefore, the initial singularity is of cigar type whenever
$\lambda\neq 0$ and of pancake type for $\lambda=0$.
This can be also inferred by means of the limit $t\rightarrow 0$
on the line-element in (\ref{eq:ds2gamma}), which after
changing $(\mu t)^\lambda \d t = \d \tau$, redefining $\mu$ and
absorbing some constants in $y$ and $z$ reads
\bean
&&\left.\d s^2\right|_{t \rightarrow 0}
=e^{-\frac{2 c \mu}{a+b} x} \left( -\d \tau^2+
(\mu \tau)^{\frac{2\lambda}{1+\lambda}} \d x^2\right)\\
&&\hspace{3cm}
+e^{-\frac{\mu}{a+b}(a+2 b l) x} (\mu \tau)^{\frac{1+2l}{1+\lambda}}\d y^2+
e^{-\frac{\mu}{a+b}(a-2 b l) x} (\mu \tau)^{\frac{1-2l}{1+\lambda}}\d z^2.
\eean
In other words,
the exponents $p_i$ such that $l_i\propto\tau^{p_i}$ \cite{Llibre} (pp. 121)
are  $\{2 p_1=2\lambda/(1+\lambda), 2 p_2=(1+2l)/(1+\lambda),
2 p_3=(1-2l)/(1+\lambda)\}$
so that
\[
p_1+p_2+p_3=p_1^2+p_2^2+p_3^2=1.
\]
Therefore, the singularity is of Kasner type \cite{lifkala,liang}.

\subsection{Future asymptotic behaviour of the solutions}
\label{sec:future}
The future asymptotic behaviour
is to be computed here as the limit $T\rightarrow \infty$ in comoving
coordinates. (The same result is obtained, of course, by
performing the limit $t\rightarrow \infty$, but the final expressions
will be eventually better related to well-known homogeneous spacetimes
by using the comoving coordinates.)  
The line-element (\ref{eq:ds2comov}) in that limit
and for finite values of the spacelike coordinate $X$ reads
\[
\left.\d s^2\right|_{T \rightarrow \infty}
=e^{2\frac{\mu}{a+b}(d-c\nu)T}\left( -\d T^2 +\d X^2\right)+
e^{\frac{\mu}{a+b}(1-\nu)(a+2bl)T}\d y^2+
e^{\frac{\mu}{a+b}(1-\nu)(a-2bl)T}\d z^2,
\]
and thus the whole family tends to the perfect-fluid self-similar
Bianchi I solutions \cite{Llibre},
the three $p_i$  exponents reading then
$p_1=1$, $p_2=(1-\nu)(a+2 b l)/(2(d-c\nu))$, 
$p_3=(1-\nu)(a-2 b l)/(2(d-c\nu))$.
The asymptotic behaviour for the
only case in which the solutions isotropize in the future
($\lambda=1/2$, $\nu=0$)
is easily identified here as the flat FLRW solution with $p=\varrho/3$
($p_1=p_2=p_3=1$).

\section{Interesting particular subfamilies}
\label{sec:subfam}
In this section we present some of the solutions included in the 
general family which are of some physical interest. We have selected 
three types of subfamilies: the dust subfamily, which was actually 
considered at length in \cite{SEVEdust} as a particular family
of a broader class of dust solutions; the $p=\varrho/3$ family, which 
has a physically realistic equation of state for radiation-dominated 
epochs and, as we will see, it provides a simple example of how a 
gravitational wave can give rise to the inhomogeneization of the 
underlying perfect fluid; and the vacuum limits, which include some 
plane-wave spacetimes and, therefore, will also be
relevant for the discussion about the inhomogeneization of the 
spacetime by the gravitational waves just mentioned. These three 
cases are treated in separate subsections.
\subsection{The vacuum limits}
\label{sec:limits}
The half-null coordinates of the previous section are very useful to 
find vacuum limits of the general solution. For the line-element as 
written in (\ref{eq:ds2hn}), the expression for the fluid velocity vector
$\vec u$ transforms to
\be
\vec u=J^{-\lambda}(u,v)
\frac{1}{\sqrt{2\ku \kv}}\left[ \ku \Paru +\kv \Parv\right],
\label{eq:vecuuv}
\ee
and the energy density reads now
$$
\varrho=J^{-2\lambda-1}\left(\frac{a-b}{a+b}\right)
\frac{(1-\nu^2)\hat{\mu}^2(\lambda+1)}
{\ku^{1+\frac{a+b}{2(d+c)}}\kv^{1+\frac{a+b}{2(d-c)}}}.
$$
 From this it follows that $\nu^2 =1$ provides vacuum limits
in this coordinate system.
Performing them in the form (\ref{eq:ds2hn}) we arrive at the 
following:
\begin{enumerate}
    \item Case with $\nu =1$. The line-element becomes
$$
ds^2=-2J^{2\lambda} du dv +J \left(J^{2l} dy^2 + J^{-2l} dz^2\right)
$$
where $J=1+\sqrt{2}\,\hat{\mu}\, v$. This is a particular pp-wave
\cite{KRAM} with $\partial_u$
as a null Killing vector. The Petrov type is N and the Weyl tensor 
takes the simple form
$$
\Psi_{0}=4 \lambda l \hat{\mu}^2 J^{-2\lambda -2}.
$$
\item Case with $\nu =-1$. Now the line-element reads
$$
ds^2=-2J^{2\lambda} du dv +\ku_{}^{\tilde{a}}J 
\left(\left[\ku^{\tilde{b}}J\right]^{2l}dy^2+
\left[\ku^{\tilde{b}}J\right]^{-2l}dz^2\right)
$$
where we have put $\tilde{a}\equiv 
2(1+4\lambda)/(1+4\lambda+16\lambda^2)$ and $\tilde{b}\equiv 
8\lambda/(1+4\lambda+16\lambda^2)$ and the function $J$ becomes
$$
J=1-\ku^{-(\tilde{a}+\tilde{b})/2}.
$$
This is again a pure gravitational pp-wave, and the type-N Weyl 
tensor has the only non-vanishing scalar
\bean
&&\Psi_{4}=-\frac{4 \lambda l \hat{\mu}^2}{J^{2\lambda +2}\ku^{\tilde{a}+
\tilde{b}+2}(8\lambda +1)}\\
&&\hspace{1cm}
\times
\left[ 1 - 4(16\lambda^2-4\lambda-1)\ku^{\tilde{a}+\tilde{b}}-
2(16\lambda+3)\ku^{(\tilde{a}+\tilde{b})/2}\right].
\eean
\end{enumerate}

The previous vacuum solutions correspond to limits of the family
of perfect fluid spacetimes of section \ref{sec:pf},
where $\nu$ was restricted by (\ref{eq:nu}).
Nevertheless, the line element given by (\ref{eq:ds2gamma})
contains further vacuum spacetimes arising whenever $a-b=0$
apart from the cases $\nu=\pm 1$, that is, if
$$
\nu\lambda +\lambda +1 =0,
$$
which together with the restriction $\lambda \geq -1/4$ requires 
$\nu >3$ or $\nu <-1$.
The line-element for these cases reads
\bea
&&\d s^2=F^2(t,x)\left(-\d t^2+\d x^2\right)+\left(
e^{\frac{\mu}{2}(t-x)}-e^{-\frac{\mu}{2}(t+x)}\right)
\nonumber\\
&&\hspace{1cm}
\times\left[\left(e^{\frac{\mu}{2}(t-x)}-e^{-\frac{\mu}{2}(t+x)}\right)^{2l}
\d y^2+
\left(e^{\frac{\mu}{2}(t-x)}-e^{-\frac{\mu}{2}(t+x)}\right)^{-2l}\d z^2\right],
\hspace{1cm}
\eea
with
\[
F(t,x)\equiv e^{-\frac{\mu}{2}x}
\left(e^{\frac{\mu}{2}(t-x)}-e^{-\frac{\mu}{2}(t+x)}\right)^{\lambda}.
\]
This vacuum subfamily has in 
general a maximal Abelian $G_{3}$ acting simply transitively
on spacelike hypersurfaces,
the additional spacelike Killing vector field given by
$$
\vec\eta=e^{\frac{\mu}{2}x}\left[\sinh\left(\frac{\mu}{2}t\right)\Part+
\cosh\left(\frac{\mu}{2}t\right)\Parx\right],
$$
and  is of Petrov type \rm{I} in general,
so that it is the well-known Kasner metric \cite{KRAM}.

\subsection{The $p=\varrho/3$ subfamily. Propagation of gravitational 
waves in a homogeneous background}
\label{sec:radia}
This section is devoted to the radiation case $\gamma=1/3$ ($\lambda=1/2$).
The condition (\ref{eq:positi}) reads then $\nu<3/7$.
As has been shown, this family constitutes a generalization
of the Ruban `non-tilted' Bianchi V solution, and we are now going to 
prove that, in fact, the general family can be interpreted as the 
inhomogeneization, via the propagation of plane gravitational waves,
of the mentioned Bianchi V Ruban solution. 

To that end, first of all it is necessary to choose the appropriate  
coordinate system in which the inhomogeneization will be more 
transparent. In this case, and despite what one might try at first, 
the natural choice is {\it not} the comoving coordinates (nor their 
half-null counterpart), but rather the original
non-comoving coordinates $\{t,x,y,z\}$. The reason for this is that 
the change to comoving coordinates (\ref{comov}) depends explicitly on the 
parameter $\nu$, but this very parameter is the one defining the 
inhomogeneization. Thus, for the several different inhomogeneous 
metrics (which are selected by the particular values of $\nu$), a 
{\it different} change to comoving coordinates is needed. In other 
words, in order to use a common coordinate system which is valid for 
the general subfamily, {\it as well as} for the particular Bianchi V metrics, 
one has to resort to the system $\{t,x,y,z\}$. This is an explicit
example of the adequacy of using non-comoving coordinates in some 
occasions based on physical grounds.

Now, we show how to write the line-element in a form that makes it 
explicit the homogeneous Bianchi V background and the travelling waves 
leading to its inhomogeneization. By setting $\lambda =1/2$, the 
metric (\ref{eq:ds2gamma}) can be rewritten as follows
\be
\d s^2=e^{f_V+f_{inh}}\left(-\d t^2+\d x^2\right)+e^{g_V+p_{inh}}
\left( e^{p_V -\sqrt{3}\, p_{inh}}\d y^2 +
e^{-p_V+\sqrt{3}\, p_{inh}}\d z^2\right),
\label{eq:ds2 1/3}
\ee
where
\be
e^{f_V}\equiv e^{\mu t} -1,\hspace{1cm}
e^{g_V}\equiv e^{f_V- \mu x},\hspace{1cm}
e^{p_V}\equiv \left(1-e^{-\mu t}\right)^{\sqrt{3}},
\label{eq:bianchiV}
\ee
correspond to the functions of the Bianchi V homogeneous solution, 
defined by $\nu =0$, whereas 
\[
f_{inh}\equiv
\mu\left(\frac{4\nu(3\nu+1)}{(1-\nu^2)(3-7\nu)}\right)(t-\nu x),
\hspace{1cm}
p_{inh}\equiv
\mu\left(\frac{4\nu}{3-7\nu}\right)(t-x),
\]
are the functions linked to the inhomogeneities of this $p=\varrho /3$
family of solutions.
The energy density for this family takes the following expression
$$
\varrho=\frac{3}{2}\;\mu^2\;
e^{-(2f_V+f_{inh})}\left(\frac{3+\nu}{3-7\nu}\right).
$$

As we see, the inhomogeneity is driven
by the same parameter $\nu$ that `tilts' the perfect fluids, that is,
the peculiar spatial velocity. This 
inhomogeneity appears primarily in the transversal part of the metric as the 
function $p_{inh}$, which depends only on the {\it null}
coordinate $t-x$,
so that $p_{inh}$ is obviosuly a solution of the
flat-space wave equation
\be
\Box_{\eta} p_{inh}=0, \label{nueva}
\ee
where $\Box_{\eta}$ denotes de d'Alembertian for the Minkowski ($\eta$)
metric. Thus,
this inhomogeneity is constant at the null hypersurfaces $t-x =$const.,
or in other words, it propagates at the speed of light
orthogonally to the surfaces of transitivity of the $G_{2}$ group of motions.
We interpret this as signalling the existence of gravitational waves 
travelling in the Bianchi V background whose spacelike propagation direction is
orthogonal to the $S_2$ orbits spanned by $\{y,z\}$.
This interpretation seems to be in accordance with that given for the
family of inhomogeneous non-diagonal stiff fluid solutions found by
Wainwright and Marshman in \cite{wainmarsh}
(see also \cite{verdaguer}), where the inhomogeneization is driven
by an arbitrary function depending on $t-x$.
In our case, the effect of the waves is a fortiori revealed also by the
imprint which leave on the longitudinal part of the line-element, 
given by $f_{inh}$. Actually, since $f_{inh}$ is functionally
dependent on the velocity potential via
$$
\sigma\propto e^{f_{inh}/4},
$$
the implicit relation between the gravitational waves
and the propagation of inhomogeneities through the acoustic waves
is manifest. The propagation of the gravitational and acoustic waves
has thus two effects: it breaks the spatial
homogeneity of the spacetime and at the same time tilts the velocity 
vector of the matter {\it but} keeping the perfect-fluid character of 
the matter content.

Of course, all this has to be considered in a more careful way. For instance,
in the case $\nu=-1/3$ we fall into the special cases analyzed in the 
final point of subsection \ref{sec:part}, so that the line-element
(\ref{eq:ds2 1/3}) is in fact a Bianchi VI$_{-3}$ spatially {\it homogeneous}
solution, but now with $p_{inh}\neq 0$ and $f_{inh}=0$ !
The form of the longitudinal part is kept, 
but there appears a non-trivial wave-like
inhomogeneity in the transversal part 
given by $p_{inh}\neq 0$. It would seem that, in this case, the 
propagation of the waves would not give rise to the inhomogeneous 
trace left in $f_{inh}$. However, this can be seen to arise as a 
rather exceptional situation because, by imposing $f_{inh}=0$ but
setting $p_{inh}=\beta (t-x)$, one obtains
a one-parameter family of Bianchi V-VI solutions for a {\it non-perfect}
fluid matter content. The third Killing vector is given by
\[
\vec \zeta=2\Parx + \left(\mu+(1-\sqrt{3})\beta\right)y\Pary+
 \left(\mu+(1+\sqrt{3})\beta\right)z\Parz.
\]
The energy-momentum tensor of this family can 
be interpreted as a fluid with a non-zero energy flux in the direction 
of propagation of the wave-like inhomogeneity. Thus, in these 
cases, the waves seem to keep the spatially homogeneous character of 
the spacetime but breaking the perfect-fluid character of the matter.
The restriction to the particular value $\beta =-\mu /4$ leads to the
mentioned case corresponding to $\nu=-1/3$, in which the matter 
content is a perfect fluid. In this exceptional case, the 
inhomogeneization effects of the travelling waves, together with that of
the flux of energy and of the acoustic waves shown by the 
tilting of the fluid, seem altogether to balance in a 
final outcome of $p_{inh}$ which simply changes the Bianchi type
of the solution.

Summarizing, one could better say it is the inhomogeneity
that generates the gravitational waves and not the other way round.
Indeed, clearly $f_{inh}\neq 0\Rightarrow p_{inh}\neq 0$, but as we 
have just seen $p_{inh}\neq 0 \not{\Rightarrow} f_{inh}\neq 0$.
This preferable point of view would state that
the tilting (or the acoustic waves) generates the inhomogeneity in $f_{inh}$
which, in turn, is responsible for the appearance of the gravitational
waves. 

The above paragraphs seriously indicate that one cannot talk 
about gravitational-wave propagation and its inhomogeneization 
properties in a naive manner. Therefore, we have tried to support the
interpretation of $p_{inh}$ representing gravitational waves propagating in 
the homogeneous background in two other independent ways. These are 
presented in what follows.

The first way is the comparison with the available systematic studies
for the description of exact gravitational waves on homogeneous backgrounds.
These studies were initially developed in \cite{Adams1}.
The exact formalism presented in that reference, later used in \cite{Adams2},
applies to perfect-fluid solutions of the Einstein field equations, 
as well as the degeneracies thereof,
containing gravitational waves propagating over
Bianchi I to VII backgrounds along an ``algebraically''
preferred direction. The case we are interested in is that of
Bianchi V cosmologies which, after inhomogeneization takes place, 
become diagonal $G_2$ perfect-fluid solutions. This case was termed as 
the single polarization (+) waves in Bianchi V, see 
\cite{Adams1,Adams2}.
Using the explicit form of the metric as given in (\ref{eq:ds2 1/3})
where $p_V$, $g_V$, $p_V$ correspond to functions of a fully
general Bianchi V spacetime and taking $p_{inh}(t,x)$ and $f_{inh}(t,x)$
to be free functions giving the inhomogeneous generalization,
the Einstein equations for a perfect fluid can be split as follows:
\be
\Box_{\eta}\left(R\;e^{-\mu x}\right)=-(\varrho-p) R\;e^{-\mu x}
\; e^{f_V+f_{inh}},
\label{eq:R}
\ee
\be
\Box \psi=0,
\label{eq:psi}
\ee
plus two other first order differential equations for $f_{inh}$
(longitudinal scale equations \cite{Adams1}).
Here $\Box$ denotes the d'Alambertian,
$R\equiv e^{f_V+p_{inh}}$ describes the transverse scale expansion and
$\psi\equiv p_V-\sqrt{3} p_{inh}$ corresponds to the so called
wave amplitude \cite{Adams1}.
It must be noticed that, in this formalism, the function
satisfying a source-free massless scalar field equation 
is $p_V-\sqrt{3} p_{inh}$, that is, the full combination
of the function corresponding to the background plus
the function carrying the inhomogeneity. This is so
for the cases with a single (+) polarization, which restricts
the Bianchi types to be I, III, V or VI, and furthermore
is in accordance with previous works in which the
diagonal Einstein-Rosen \cite{einsrosen} form of the line-element
is taken and $\psi$ is the function appearing in the transverse part once the
transitivity surface area element, which corresponds to
$R\;e^{-\mu x}$ in (\ref{eq:ds2 1/3}),
is factorized, see \cite{liang,CCM,verdaguer,bicakgrif}. 
Nevertheless, as we see in the previous case, the transverse
scale expansion $R$ as defined in \cite{Adams1} {\em is not} equivalent
to the transitivity surface area element in general.
Actually, for some other Bianchi types this implies
that the definition of ``$\psi$'' in \cite{Adams1,Adams2} differs 
from that in the ``Einstein-Rosen view'', and thereby
the wave equation they satisfy are different.

The function $p_{inh}(t,x)$ will
actually satisfy the homogeneous wave equation $\Box p_{inh}=0$
for those cases with $\varrho=p$, that is, for stiff fluids
including the vacuum and the minimally coupled massless
scalar field solutions. This is so primarily because the so-called
``transverse scale'' equation (\ref{eq:R}) couples $R$
to the matter source through
the function $\varrho-p$, so as long
as this function vanishes, $R \;e^{-\mu x}$
is completely independent of the matter content
and satisfies a flat-space wave equation. One can argue then
that $R$ has to be in fact the ``homogenous'' function $R=R_V(t)$ in order
to have models that reduce to the Bianchi spacetime when the waves
reduce to zero, see \cite{Adams2}.
In other occasions, as was the case in the ``Einstein-Rosen view'',
the condition $R_{,\mu} R^{,\mu}<0$ everywhere was often imposed,
see \cite{liang}.
These simplifications, for the stiff fluid case, imply eventually that
the d'Alambertian $\Box$ coincides with $\Box_V$, that is, with the
d'Alambertian of the homogeneous background metric,\footnote{We are
concentrating in the Bianchi V case, but similar
scenarios arise in the rest of the Bianchi I to VII spacetimes \cite{Adams1}.}
and thus, since $\Box p_V=\Box_V p_V=0$, equation (\ref{eq:psi}) 
implies $\Box p_{inh}=0$. That is, both $p_V$ and $p_{inh}$ are solutions
of the same wave equation (\ref{eq:psi}).

Fortunately, despite all the above, the simplification on $R$ is not necessary
and, moreover, for more realistic equations of state $R$ evolves
(as must be!) {\it coupled} with the matter. The former statement follows 
trivially in general because, in (\ref{eq:ds2 1/3}), the homogeneous spacetime
is recovered when $p_{inh}=0$, while the latter immediately implies 
that, in fact, $\Box \neq \Box_V$. In (\ref{eq:ds2 1/3}) they are related
as follows
$$
\Box=e^{-f_{inh}}\left[\Box_V -\dot{p}_{inh} \partial_t+
{p'}_{inh} \partial_x\right],
$$
and therefore the coupling of $R$ with the matter leads to the appearance
of $p_{inh}$, causing the inhomogeneization of the
transverse scale expansion, which in turn drives the coupling of the
function $\psi$ with the matter through the operator $\Box$.
The longitudinal scale equations would account then for $f_{inh}$.
Notice that, as remarked before, in the present family (\ref{eq:ds2 1/3}) with
(\ref{eq:bianchiV}), it is $f_{inh}$, or equivalently $\sigma$,
that actually switches on the inhomogeneization.
In (\ref{eq:ds2 1/3}) together with (\ref{eq:bianchiV}) and because
$p_{inh}$ is a function of $t-x$, we have that
$$
\Box p_{inh}=e^{-f_{inh}} \Box_V p_{inh}=\Box p_V/\sqrt{3}=
-e^{-(f_{inh}+f_V)}\frac{\mu}{\sqrt{3}} \left(\frac{4\nu}{3-7\nu}\right)
\dot{p}_V,
$$
the second  equality coming from (\ref{eq:psi}). Thus, the
propagation of $p_{inh}$ is not that of a {\it source-free}
massless scalar field, rather it is driven by the interaction with
the background geometry transverse part,
clearly showing the non-linearity on the splitting of the background
and the wave. The important thing here, however, is that the above 
equation is manifestly hyperbolic in character, as is obvious in its 
simpler form (\ref{nueva}), and that the characteristic propagation 
speed of its solutions is the speed of light. Having no 
electromagnetic field present, the propagation of gravitational waves 
seems the best possibility.

The second way to support our claim comes from the vacuum limits 
of the solutions presented in subsection \ref{sec:limits}.
Due to the non-linear interaction
of the waves with the matter proper to General Relativity,
and as we have seen in the previous paragraph for this particular case,
it is not  possible nor desirable to separate the metric into two
linear terms representing the background and the waves respectively.
However, one can try to annihilate completely one of the two terms,
and then only 
the other must survive. Of course, by construction, if we set
the wave-like part defined by $p_{inh}$ to zero we obtain the Bianchi 
V, $p=\varrho /3$, homogeneous background. This corresponds to putting $\nu=0$.
But can we also get rid of the matter, so that only the gravitational 
wave remains? A striking and beautiful answer would come from
finding that the vacuum limits in the coordinates of (\ref{eq:ds2 1/3})
correspond to plane gravitational waves, as was the case for the 
limits found in subsection \ref{sec:limits}. Unfortunately, there are
no vacuum limits in these coordinates.
To achieve the vacuum limits one has to make use of other
coordinate systems, but then again
the limit depends on the coordinate system chosen \cite{limitsge}.
In spite of this, the existence of coordinate systems in which the
limits $\nu=\pm 1$ correspond to a gravitational pp-wave
has already been shown in subsection \ref{sec:limits}.
At this point, the question giving unequivocal sense to the
gravitational wave inhomogeneity interpretation for this
family of $p=\varrho/3$ solutions would thus be: do all the vacuum limits
of this family with $\nu=- 1$ correspond to a plane gravitational wave?
We do not have a rigorous answer for this question yet, but we do believe 
that the answer is positive. In this sense, we claim that getting rid 
of the matter, {\it whenever this is possible}, and within the allowed 
range of the parameter $\nu$, provides a pure and simple gravitational 
pp-wave, which we interpret as the remanent of the mixed case.

Hitherto, the work on gravitational waves in cosmological backgrounds
based on exact solutions has been mainly tackled
in the cases of vacuum, scalar and electromagnetic fields, and
stiff fluid (and its anisotropic generalization \cite{letelier1})
as sources in $G_2$ on $S_2$ spacetimes, 
see the reviews \cite{CCM,Adams1,verdaguer,bicakgrif} and references therein.
As mentioned in the Introduction,
there are other works presenting solitonic perturbations of
$p=\varrho /3$ FLRW spacetimes \cite{ibver}, later generalized to
$p=\gamma \varrho $ in \cite{diglepul}, but
the energy-momentum tensor of the inhomogeneous spacetime
turns out to be a non-perfect fluid.
The general formalism used in these papers
assumes that the backgrounds in which the perturbations
propagate have a restricted type of
energy-momentum tensors, so it has been applied only to FLRW spacetimes.
The present family of solutions may constitute then the first
exact solution in General Relativity for a perfect fluid with a
realistic $p=\varrho /3$
equation of state describing gravitational waves travelling on a 
spatially homogeneous background.

\subsection{The dust subfamily}
\label{sec:dust}
The particular dust cases, defined by $\lambda=0$ (so that (\ref{eq:positi})
requires now $\nu<0$), belong to
a more general family of algebraically general
dust spacetimes already presented in \cite{SEVEdust}.
We devote this short subsection to identifying the present dust
one-parameter subfamily within the more general dust family appearing
in \cite{SEVEdust}.

By taking profit of the form
of the line-element in the half-null
coordinates $\{u,v,y,z\}$ (\ref{eq:ds2hn}), and by
making a coordinate change to Lorentzian
coordinates analogous to that in (\ref{eq:UV}),
the line-element (\ref{eq:ds2hn}) with $\lambda=0$
coincides exactly with expression (7) of \cite{SEVEdust} for the following
particular values of the constants (here, we denote with a tilde the 
quantities which appear in reference \cite{SEVEdust}, if necessary):
\begin{itemize}
    \item $\tilde{a}=\kappa_u/\sqrt{2}$. This can be seen as the free 
    parameter of the solution.
    \item $c_2=0$, $c_1<0$. This implies the relation $\kappa_u 
    =\kappa_v$, which follows from $\lambda =0$.
    \item $\tilde{b}=\nu(1+\nu)/(1+\nu^2)$. By remembering that
    $\nu\in (-1,0)$, this leads to the restriction
    $\tilde{b}\in [\tilde{b}_-,0)$ for the parameter $\tilde{b}$ of
    \cite{SEVEdust}.
\end{itemize}
The diagram for this subfamily corresponds to the
figure 2(f) of \cite{SEVEdust} with the point $\tilde{p}$ at
$\tilde{t}=\tilde{x}=0$,
in agreement with the particular case (c)
of the present figure \ref{fig:main}.

\section*{Acknowledgements}
We are grateful to Alex Feinstein, Filipe C. Mena and Reza Tavakol
for some fruitful comments and helpful indications concerning references.
The authors also thank financial support from the Basque Country 
University, from the Generalitat de Catalunya, and from the Ministerio 
de Educaci\'on y Cultura, under project numbers UPV 172.310-G02/99,
98SGR 00015,
and PB96-0384, respectively. RV thanks the
Spanish Secretar\'{\i}a de Estado
de Universidades, Investigaci\'on y Desarrollo,
Ministerio de Educaci\'on y Cultura, grant No. EX99 52155527.


\begin{thebibliography}{99}
\bibitem{Llibre}Wainwright J and Ellis G F R, eds. (1997)
{\it Dynamical Systems in Cosmology\/},
Cambridge University Press, Cambridge
\bibitem{KRAM}Kramer D, Stephani H, Herlt E, and MacCallum M A H (1980)
{\em Exact solutions of Einstein's field equations}, Cambridge
University Press, Cambridge 
\bibitem{totxosing}Senovilla J M M \Journal{\GRG}{30}
{701}{-848}{1998}
{Singularity theorems and their consequences}
\bibitem{llibrePetrov}Petrov A Z 1966 {\em New methods in general relativity}
(in russian), Nauka, Moskow. [English edition: {\em Einstein spaces},
Pergamon Press, New York (1969)] 
\bibitem{ryan}Ryan M P and Shepley L C (1975) {\it Homogeneous relativistic
cosmologies\/}, Princeton University Press, Princeton
\bibitem{KR}Krasi\'nski A (1997) {\em Inhomogeneus Cosmological Models},
Cambridge University Press, Cambridge 
\bibitem{WW1}Wainwright J \Journal{\JPA}{14}
{1131}{-1147}{1981}
{Exact spatially inhomogeneous cosmologies}
\bibitem{alitesi}Sintes A (1997) {\it Inhomogeneous cosmologies
with special properties\/}, {\it Ph. D. Thesis}
Universitat de les Illes Balears
\bibitem{Raultesi}Vera R (1998) {\it Theoretical aspects concerning
separability, matching and matter contents of inhomogeneities
in cosmology\/},
{\it Ph. D. Thesis} Universitat de Barcelona
\bibitem{ali}Carot J and Sintes A M \Journal{\CQG}{14}
{1183}{-1205}{1997}
{Homothetic perfect fluid spacetimes}
\bibitem{MATHO}Mars M and Wolf T \Journal{\CQG}{14}
{2303}{-2330}{1997}
{$G_2$ perfect-fluid cosmologies with a proper conformal Killing vector}
\bibitem{HEWWAIN}Hewitt C G and Wainwright J \Journal{\CQG}{7}
{2295}{-2316}{1990}
{Orthogonally transitive $G_2$ cosmologies}
\bibitem{RUSE}Ruiz E and Senovilla J M M \Journal{\PRD}{45}
{1995}{-2005}{1992}
{General class of inhomogeneous perfect-fluid solutions}
\bibitem{WAGO}Wainwright J and Goode S W \Journal{\PRD}{22}
{1906}{-1909}{1980}
{Some exact inhomogeneous cosmologies with equation of state $p=\gamma
\varrho$}
\bibitem{AGGO}Agnew A F and Goode S W \Journal{\CQG}{11}
{1725}{-1742}{1994}
{The $p=\mu$ separable $G_2$ cosmologies with heat flow}
\bibitem{G2NC}Senovilla J M M and Vera R \Journal{\CQG}{15}
{1737}{-1758}{1998}
{$G_2$ cosmological models separable in non-comoving coordinates}
\bibitem{EM}Ellis G F R and MacCallum M A H \Journal{\CMP}{12}
{108}{-141}{1969}
{A class of homogeneous cosmological models}
\bibitem{ruban}Ruban V A \Journal{\SPJ}{45}
{629}{-637}{1977}
{Dynamics of anisotropic homogeneous generalizations of the Friedmann
cosmological models}
\bibitem{SEVEdust}Senovilla J M M and Vera R \Journal{\CQG}{14}
{3481}{-3487}{1997}
{Dust $G_2$ cosmological models}
\bibitem{alexjose1}Feinstein A and Senovilla J M M \Journal{\CQG}{6}
{L89}{-L91}{1989}
{A new inhomogeneous cosmological perfect fluid solution with
$p=\varrho/3$}
\bibitem{davidson90}Davidson W \Journal{\JMP}{32}
{1560}{-1561}{1990}
{A big-bang cylindrically symmetric radiation universe}
\bibitem{senosing}Senovilla J M M \Journal{\PRL}{64}
{2219}{-2221}{1990}
{New class of inhomogeneous cosmological perfect fluid solutions without
big-bang singularity}
\bibitem{Marctesi}Mars M (1995) {\it Geometric
properties of hypersurfaces and axial symmetry with applications
to $G_2$ perfect-fluid solutions\/},
{\it Ph. D. Thesis} Universitat de Barcelona
\bibitem{MASEbi}Mars M and Senovilla J M M \Journal{\CQG}{14}
{205}{-226}{1997}
{Non-diagonal $G_2$ separable perfect-fluid spacetimes}
\bibitem{lifkala}Lifshitz E M and Khalatnikov I M \Journal
{\it Usp. Fiz. Nauk}{80}{391}{}{1963}{}: \Journal{\it Sov. Phys. Uspekhi}{6}
{495}{-522}{1964}
{Problems of relativistic cosmology}
\bibitem{liang}Liang E P \Journal{\AJ}{204}
{235}{-250}{1976}
{Dynamics of primordial inhomogeneities in model universes}
\bibitem{wainmarsh}Wainwright J and Marshman B J \Journal{\PPLL}{72}
{275}{-276}{1979}
{Some exact cosmological models with gravitational waves}
\bibitem{CCM}Carmeli M, Charach Ch and Malin S \Journal{\it Phys. Rep.}
{76}{79}{-156}{1981}
{Survey of cosmological models with gravitational, scalar and
electromagnetic waves}
\bibitem{Adams1}Adams P J, Hellings R W, Zimmerman R L, Farhoosh H,
Levine D I and Zeldich S \Journal{\AP}{253}
{1}{-18}{1982}
{Inhomogeneous cosmology: gravitational radiation in Bianchi backgrounds}
\bibitem{verdaguer}Verdaguer E \Journal{\PREP}{229}
{1}{-80}{1993}
{Soliton solutions in spacetimes with two spacelike Killing fields}
\bibitem{bicakgrif}Bic\'ak J and Griffiths J B \Journal{\ANY}{252}
{180}{-210}{1996}
{Gravitational waves propagating into Friedmann-Robertson-Walker Universes}
\bibitem{letelier1}Letelier P S \Journal{\PRD}{26}
{2623}{-2631}{1982}
{Solitary waves of matter in general relativity}
\bibitem{alekseev}Alekseev G A \Journal{\it Proc. Stecklov Inst. Math}{3}
{215}{-262}{1988}
{Exact solutions in the general theory of relativity}
\bibitem{ibver}Ib\'a\~nez J and Verdaguer E \Journal{\AJ}{306}
{401}{-410}{1986}
{Finite perturbations on Friedmann-Robertson-Walker models}
\bibitem{diglepul}D\'{\i}az M C, Gleiser R J, Pullin J A \Journal{\CQG}{4}
{L23}{-L28}{1987}
{Solitonic perturbations of perfect fluid Friedmann-Robertson-Walker
cosmological models}
\bibitem{SteEll} Stewart J M and Ellis G F R \Journal{\JMP}{9}
{1072}{-1082}{1974}
{Solutions of Einstein's equations for a fluid which exhibit local
rotational symmetry} 
\bibitem{henkEll} van Elst H and Ellis G F R \Journal{\CQG}{13}
{1099}{-1127}{1996}
{The covariant approach to LRS perfect fluid spacetime geometries}
\bibitem{UgRo} Uggla C and Rosquist K \Journal{\CQG}{7}
{L279}{-L283}{1990}
{New exact perfect fluid solutions of Einstein's equations II}
\bibitem{limitsge}Geroch R \Journal{\CMP}{13}
{180}{-193}{1969}
{Limits of Spacetimes}
\bibitem{Adams2}Adams P J, Hellings R W, Zimmerman R L \Journal{\AP}{288}
{14}{-21}{1985}
{Inhomogeneous Cosmology. II. Linearly Polarized Gravitational Waves}
\bibitem{einsrosen}Einstein A and Rosen N \Journal{\it J. Franklin Inst.}{223}
{43}{}{1937}{On gravitational waves}





\end{thebibliography}
\end{document}